# Epitaxial checkerboard arrangement of nanorods in ZnMnGaO4 films studied by x-ray diffraction

S. M. O'Malley,[a] P. L. Bonanno, K. H. Ahn, and A. A. Sirenko

Department of Physics, New Jersey Institute of Technology, Newark, New Jersey 07102

A. Kazimirov

Cornell High Energy Synchrotron Source (CHESS), Cornell University, Ithaca, New York 14853

S. Park and S-W. Cheong

Rutgers Center for Emergent Materials and Department of Physics and Astronomy, Rutgers University, Piscataway, New Jersey 08854

---

[a] so26@njit.edu

# ABSTRACT

The intriguing nano-structural properties of a ZnMnGaO$_4$ film epitaxially grown on MgO (001) substrate have been investigated using synchrotron radiation-based x-ray diffraction. The ZnMnGaO$_4$ film consisted of a self-assembled checkerboard (CB) structure with perfectly aligned and regularly spaced vertical nanorods. The lattice parameters of the orthorhombic and rotated tetragonal phases of the CB structure were analyzed using H-K, H-L, and K-L cross sections of the reciprocal space maps measured around various symmetric and asymmetric reflections of the spinel structure. We demonstrate that the symmetry of atomic displacements at the phases boundaries provides the means for coherent coexistence of two domains types within the volume of the film.

## I. INTRODUCTION

The precise control of domains and phase-separation on the nanoscale can provide new technological advantages for monolithic integration of materials with complementary electronic and magnetic properties. Recent interest in the growth techniques that are not restricted by the spatial resolution of lithography, but based on self-organization in oxides, has resulted in discoveries of several materials with the periodic pattern of nano-checkerboard (CB) domains [1,2,3,4,5,6,7]. Synthesis of spinel oxides with CB structures, such as nonmagnetic $ZnMnGaO_4$ (ZMGO) [1] and magnetic Mn-doped $CoFe_2O_4$ [4], relies on harnessing Jahn-Teller (JT) distortions in the solid-phase during the growth. For example, the self-assembled CB structure in bulk polycrystalline ZMGO [1] occurs during the annealing process: upon cooling the high temperature cubic ZMGO phase experiences a JT distortion of its cubic phase into a lower symmetry tetragonal phase, inducing a miscibility gap within the system of mixed JT ions and subsequent spinoidal decomposition. The mixture then experiences a diffusion-type spatial separation into two phases, Mn-rich and Mn–poor, regions, driven by the aggregative nature of JT-ions to form clustered regions. Interaction between the alternating Mn-rich (tetragonal) and Mn-poor (cubic) regions can then coexist by producing the strain-accommodating CB structure [1,3,4,8,9]. An important step towards utilization of this intriguing phenomenon for the large-scale planar device technology have been recently reported in Ref. [4], where thin films of $ZnMnGaO_4$ with perfectly aligned nanorods with a CB pattern were epitaxially grown on MgO substrates. In this paper we discuss structural properties of these $ZnMnGaO_4$ films, such as lattice parameters, relaxation effects, rotation angles of nanodomains, and atomic displacements at the domain boundaries.

## II. EXPERIMENT

The ZMGO films with the thickness of 820 nm were grown by pulsed laser disposition (PLD) on a single crystal MgO (001) substrate with a cubic structure ($a = 0.4212$ nm) using a KF excimer laser with a repetition rate of 5 Hz and an energy density of 1.0 J/cm$^2$. The MgO substrate was heated to 570 $^o$C with a 300 mTorr oxygen partial pressure; after deposition the sample was annealed for ~8 hours in an 400 Torr oxygen rich environment. The target material consisted of a homogenous high temperature quenched form of ZMGO. The CB structure is formed by two periodically alternating and structurally different spinel phases: (i) rotated tetragonal (Mn-poor, JT inactive) and (ii) orthorhombic phases (Mn-rich, JT-active). The CB structure consists of regularly spaced nanodomains with the typical size of 4×4×750 nm$^3$, which are aligned along the film growth direction. More growth details have been described in Ref. [4].

X-ray diffraction measurements were carried out at the A2 beamline of Cornell High Energy Synchrotron Source (CHESS). The incident synchrotron beam was conditioned using a double-bounce Si (111) monochromator with the passing energy of 10.53 keV and a pair of orthogonal slits determining the 0.2×0.2 mm$^2$ beam size. Sample was mounted in a four-circle Huber diffractometer. High angular resolution was achieved by conditioning the diffracted beam with a single-bounce Si (111) analyzer crystal, before being collected by a Bede scintillation detector. Reciprocal space mapping (RSM) was facilitated by the SPEC software for the movement of the diffractometer along arbitrary (H K L) vectors in the reciprocal space. The variable diffraction angles θ, 2θ, and χ allowed measurements of the reciprocal space cross sections H–L, K–L, and H–K in the regime of a fixed ϕ angle (ϕ corresponds to rotation around the sample normal direction). RSMs for the CB structures were analyzed for several symmetric and asymmetric reflections: (0 2 2), (2 2 2), (0 0 4), (0 4 4), (2 2 6), (0 2 6), and (6 6 6). In the following, Miller index L corresponds to the film growth direction, while H and K represent the in-plane parameters of the structure. The integer values of H, K, and L correspond to the reciprocal lattice points of a reference spinel structure with the lattice parameter $a_0$ = 0.8424 nm (twice the value of that for the cubic MgO substrate). In this notation a strong (0 2 2) diffraction peak of MgO substrate coincides with the (0 4 4) reciprocal lattice point (RLP) of the reference spinel structure.

Performing a series of symmetric and asymmetric cross-sectional RSMs has allowed for the determination of both in- and out-plane lattice parameters along with the presence of structural distortions within the CB structure. The H–K, H–L, and K–L cross sections around the (0 4 4) and (0 0 4) RLPs are compared in Figure 1(a,b). The H–K map for the (0 4 4) reflection is dominated by several peaks, which correspond to different phases within the CB film. In Ref. [4] we attributed the following four broad peaks labeled α, β, γ, and δ to two rotated tetragonal (β and γ) and two perpendicularly-oriented orthorhombic (α and δ) phases. The L value for the centers of α, β, γ, and δ peaks is the same and is equal to 4.07 [see side panels in Fig. 1(a)]. This observation indicated that all α, β, γ, and δ phases have the same out-plane lattice parameter, which is crucial for their coexistence in the volume of the film. The ΔK/K values for two orthorhombic phases are 0.030 for α and −0.040 for δ phases. The average in-plane lattice parameter of the two rotated tetragonal phases is practically lattice-matched to the substrate. The centers of each tetragonal peaks (β and γ) are rotated by ±2.55° around the (0 0 L) reciprocal lattice vector. The assignment of the rotation axis is based on comparison with the RSMs measured at the symmetric (0 0 4) reflection. In Fig. 1(b), none of the α, β, γ, and δ peaks appear at the same (H, K) positions as that for the (0 4 4) reflection [10]. In contrast, the H–K cross section for (0 0 4) reflection shows that all of the α, β, γ, and δ peaks have collapsed into a single peak around H=K=0, while the corresponding intensity maximum has the same value of L = 4.07 as that for the (0 4 4) reflection [see side panels in Fig. 1(b)].

The full width at the half maximum (FWHM) values in the H–K plane $\xi_{H-K}$ were determined for both the (0 4 4) and (0 2 2) reflections. Contributions due to mosaic spread, *i. e.* the tilt and twist variations, along with the coherent domain size were decomposed from the peaks FWHMs through their dependence on the reflection order [11]. The in-plane correlation length $D_{H-K}$ for the CB domains has been estimated to be about 25 nm, where $D_{H-K} = a_0/\xi_w$ and $\xi_w$ is the reflection-order-independent component of $\xi_{H-K}$. The tilt distortion is believe to be negligible, because the orientation of the CB peaks in the H–L and K–L planes remains parallel to the sample surface. Note that $D_{H-K} \approx 25$ nm is significantly larger than the typical size of the CB square of ~ 4×4 nm² (see TEM in Ref. [4]). In the L direction, the corresponding FWHM is significantly narrower $\xi_L = 0.03$ thus corresponding to inhomogeneous broadening due to composition fluctuations in the CB structure. The experimental values of the lattice parameters of the CB phases are summarized in Table 1.

The sharp peak $A$ at the center of the H–K map in Fig. 1(a) originates from the tetragonal phase that is lattice-matched to the substrate. The strain in this layer is elastic as it is evident from the (0 4 4), (2 2 6), and other asymmetric reflections. The relative position of $A$ and $\alpha$, $\beta$, $\gamma$, and $\delta$ peaks in the H–L and K–L planes can be seen in the side panels of Fig. 1(a). The corresponding L value for $A$ phase is equal to 4.095 ($\Delta L_A/L = 0.023$) that is more than that for $\alpha$, $\beta$, $\gamma$, and $\delta$ phases ($\Delta L_{CB}/L = 0.016$). Intensity of each $\alpha$, $\beta$, $\gamma$, and $\delta$ peaks for (0 4 4) reflection has been integrated over the volume of the reciprocal space and it turns out to be the same within 10%. The total diffracted intensity for these four phases, however, is 6 times more than that for a much narrower peak $A$. Comparison of the relative diffracted intensity and the lattice mismatch in L direction supports the assignment of peak $A$ to a thin transitional layer between the MgO substrate and the CB structure.

Figure 2 shows H–K RSMs measured around the (-2 -2 2), (0 2 2), and (-2 0 2) RLPs with exact value of L=2.04. The similarity between the (-2 0 2) and (0 2 2) reflections corresponds to a 90° rotational invariance (with translation) of the entire CB structure. All diffraction peaks for the (-2 2 2) reflection are closely aligned to the arc with a constant magnitude of the reciprocal lattice vector with the radius of $R = 2\sqrt{2}$ shown with a dashed line in Figure 2(a). This observation corresponds to a close proximity of the *average* in-plane lattice parameter for all $\alpha$, $\beta$, $\gamma$, and $\delta$ phases to that for the reference spinel structure, and, hence, the average in-plane parameter of the CB structure matches that for the MgO lattice. Orthorhombic and rotated tetragonal phases are separated by the domain boundaries (DB) closely aligned along the [1 1 0] and [1 -1 0] directions as shown in Figure 2(d). These domain walls should accommodate structural distortions between α, β, γ, and δ phases providing means for their coherence along both the film growth direction and in-plane direction. From comparison of the in-plane domain size (4×4 nm$^2$) and the in-plane footprint of the spinel lattice (~ 0.8×0.8 nm$^2$), the fraction of the distorted unit cells at the DB compared to the number of undistorted cells inside each phase is significant: about 30%. Hence, these domain walls should also produce a significant contribution to the total diffraction picture. Thus, we assign the DB diffraction signal to four additional diagonal streaks labeled ρ, σ, τ, and υ and positioned between α, β, γ, and δ peaks in Figs 1(a), 2(b), and 2(c). Figure 2(d) illustrates the CB arrangement of tetragonal and orthorhombic domains and the DB along the [1 1 0] and [1 -1 0] directions. The dashed lines define the CB super-cell by which translational operations can repeat the entire CB film. The angles between the DB have been determined from the angular separation between ρ, σ, τ, and υ streaks in the H-K maps. The 90° angle between the υ–τ and ρ–σ streaks corresponds to domains with the twisted tetragonal phases, while the 97.5° and 82.5° separation of the ρ–τ, and σ–υ streaks confirms presence of the twined, (perpendicularly oriented) rhombus-shaped domains with the orthorhombic phases. The $\Pi - L$ cross section of the reciprocal space taken along these diagonal streaks, where $\Pi = (H + K)/\sqrt{2}$, is shown in Fig. 3(b) for the (-2 0 2) reflection, and compared with the K-L cross section shown in Fig. 3(a). As expected for coherent DB peaks, the out-plane lattice parameters for ρ, σ, τ and υ is the same as that for α, β, γ, and δ peaks.

## III. DISCUSSION

To better understand the symmetry of the lattice distortions at the domain boundaries, it is useful to describe the *in-plane* CB pattern in terms of distortions with respect to a 2D square lattice. The mode-based atomic-scale description of the lattice distortions, reminiscent of Ginzburg-Landau approach, has been recently developed for a 2D square lattice with a monatomic basis [12,13]. Six modes relevant to the CB pattern are shown in Figure 4(a). Mode $e_3$ corresponds to square-to-rectangle distortions with $a/b$ ratio $(1+e_3/\sqrt{2})/(1-e_3/\sqrt{2})$. If we represent $e_3 = \varepsilon > 0$ in the domain $\alpha$, the domain $\delta$, has $e_3 = -\varepsilon$. Mode $r$ represents rotations with the rotation angle given by $r/\sqrt{2}$ radian. The amplitude of the mode $r$ is $r = \varepsilon$ in the domain $\gamma$ and $r = -\varepsilon$ in the domain $\beta$. The distortions at the DBs can be represented as a linear combination of $e_3$, $r$, and two other modes, either $t_+$ and $s_+$ or $t_-$ and $s_-$. For example, the distortion at the interface between $\gamma$ and $\delta$ domains in Fig. 4(b) has the character of $-e_3 + r + t_+ + s_+$, with distortions of $e_3 = -\varepsilon/2$, $r = \varepsilon/2$, $t_+ = \varepsilon/2$, and $s_+ = \varepsilon/2$. Among these four modes, $e_3$ and $r$ are the strain modes, repetition of which can fill the entire space and shift the corresponding diffraction peak around the undisturbed RLP of the reference lattice. In contrast, $s_+$ is a short wavelength $(\pi, \pi)$ mode that should alternate its sign and double the unit cell to fill the entire space. Such short wavelength distortions could result in redistribution of the x-ray intensity between different reflection orders and could generate super-lattice peaks. Mode $t_+$ represents a rigid translation, the application of which does not change the x-ray diffraction pattern. Therefore, the distortions $e_3 = -\varepsilon/2$ and $r = \varepsilon/2$ within the DB could shift the diffraction peak from the undisturbed RLP position to the point right between the peaks from domain $\delta$ and domain $\gamma$, which corresponds to the peak $\tau$ in Figs. 1 and 2.

If we consider the scattering only from a single segment of this interface and neglect the effect of the interference with the rest of the sample, the FWHM of the τ peak in the H–K plane should be inversely proportional to the width of the DB, and therefore should be very broad overlapping with δ and γ peaks. The relatively narrow FWHM of the τ peak, as well as other ρ, σ, and υ peaks, is an experimental evidence of highly coherent lattice distortions among the DB and the neighboring domains or among interfaces of the same kind, the interference effect from which can modulate the total diffraction intensity. Note that this observation is in agreement with the aforementioned large value of the correlation length for all CB domains: $D_{H-K} \approx 25$ nm that covers several identical DBs between 4×4 nm$^2$ CB domains.

Thus, XRD results allow us to reconstruct structural properties of the CB film. The MgO substrate is covered by a thin elastically-strained ZnGaMnO$_4$ layer with no phase separation (peak *A*), where the relaxation process begins as evident in the tail of peak *A* [Fig. 1(a)] between $\Delta L_A/L = 0.023$ and $\Delta L_{CB}/L = 0.017$. Eventually, accumulation of the volume strain energy results in partial relaxation of strain and formation of the elastically-strained CB layer consisting of two conversely rotated tetragonal and two orthogonal orthorhombic phases (α, β, γ, and δ). The corners of the CB super cell match perfectly the MgO substrate. The DBs, which are oriented close to [1 1 0] and [1 -1 0] directions, separate orthorhombic and rotated tetragonal phases and accommodate the structural imparity between α, β, γ, and δ phases by means of the lattice distortions shown in Fig. 4(b). The accommodating mechanism is determined by the contribution of Mn$^{3+}$ ions, which become JT-active upon occupying the octahedral sites in the spinel structure of the Mn-rich orthorhombic phase. The orbital degeneracy of the Mn ions ground state is lifted by structural distortions and the electric energy gain overcomes the cost of displacive structural energy.

## IV. CONCLUSIONS

Among open questions for the future theoretical analysis of the CB structures we mention modeling of the exact in-plane size of the pattern and the mechanisms of suppression of the herring-bone structure for films *vs.* that in the bulk. Here we will only speculate that the size of the CB domains is determined by various competing factors. The energy cost for the DBs would favor bigger domains since the relative number of the strained lattice cells at the DBs to the number of undisturbed cells decreases inversely proportional to the domain size. However, the domain size is limited from above by the diffusion length of Mn and Ga ions during the annealing, as indicated in a recent experiment on nano CB formation of Mn-doped $CoFe_2O_4$ spinel compound [4]. Other factors, such as the strain energy cost between the CB film and the substrate, would also influence the size of domains, giving rise to a nanometer length scale CB pattern and suppressing the herring-bone structure formation.

## ACKNOWLEDGEMENTS

Work at NJIT and Rutgers was supported by the DE-FG02-07ER46382 and the NSF-DMR-0546985. Work at The Cornell High Energy Synchrotron Source is supported by the NSF and the NIH/NIGMS under Award No. DMR-0225180. Authors are thankful to Theo Siegrist for interest and useful discussions.

**TABLE 1.** Average lattice parameters and related H-K-L valves of different phases of the CB film are determined from multiple reflections. The lattice and volume mismatch for different phases is normalized to that for the reference spinel structure (sp) with $a_0 = 0.8424$ nm.

|  | ZnMnGaO$_4$ [4] (bulk, $I4_1/amd$) | A | $\alpha$ / $\delta$ | $\beta$, $\gamma$ |
|---|---|---|---|---|
| $\Delta H/H$ | ___ | 0.00 | −0.03 / 0.04 | 0.00 |
| $\Delta K/K$ | ___ | 0.00 | 0.04 / −0.03 | 0.00 |
| $\Delta L/L$ | ___ | 0.023 | 0.016 | 0.016 |
| $a$, nm | 0.82 | 0.8424 | 0.898 / 0.814 | 0.841 |
| $b$, nm | 0.82 | 0.8424 | 0.814 / 0.898 | 0.841 |
| $c$, nm | 0.87 | 0.823 | 0.829 | 0.829 |
| $V/V_{sp}$ | −0.022 | −0.023 | +0.013 | −0.020 |
| $\xi_{H-K}$ (0 4 4) | ___ | 0.01 | 0.13 | 0.13 |
| $\xi_L$ (0 4 4) | ___ | 0.02 | 0.03 | 0.03 |
| Rotation [0 0 L] | ___ | 0.00 | 0.00 | 2.55° |

## FIGIRE CAPTIONS:

**FIG. 1.** (color online). (a) The H–K, H–L, and K–L cross sections of the reciprocal space measured around the asymmetric (0 4 4) reflection. For H–K map L = 4.08, $\alpha$ and $\delta$ are orthorhombic CB phases, $\beta$ and $\gamma$ are rotated tetragonal ones. The elastically strained tetragonal phase is labeled $A$. The diffracted signal from domain boundaries is labeled $\rho$, $\sigma$, $\tau$ and $\upsilon$. The H–L cross section in (a) illustrates the difference between $\Delta L_{CB}$ and $\Delta L_A$. Solid lines depict directions from (0 0 4) to $\beta$ and $\gamma$. Inset in the center is a 3D reconstruction of the diffracted intensity where the ellipse size is equal to the experimentally determined FWHMs of the corresponding peaks. (b) RSMs for symmetric (0 0 4) reflection.

**FIG. 2.** (color online). (a,b,c) The H–K cross sections (L = 2.04) of the reciprocal space around the (-2 -2 2), (0 2 2), and (-2 0 2) RLPs. For illustrative purposes the (-2 -2 2) map is converted to depict the (-2 2 2) reflection. Calculated positions of the diffraction peaks for $\alpha$, $\beta$, $\gamma$, and $\delta$ phases are marked with open triangles, circles, and stars. Solid lines depict directions from the (0 0 2) RLP towards the rotated tetragonal peaks $\beta$ and $\gamma$. Dashed ark represents the direction of rotation for two tetragonal phases. (d) CB structure phases with atomic distortions at the DBs. The super cell of the CB structure is highlighted with dashed contour.

**FIG. 3.** (color online). (a) The K–L cross sections of the reciprocal space around (-2 0 2) RLP. (b) The $\Pi - L$ cross section for the same reflection along the diagonal

streaks. Note that the L value for σ and υ peaks is the same as that for α, β, γ, and δ peaks in the H–L, and K–L cross sections.

FIG. 4. (color online). (a) The normal distortion modes of a monoatomic 2D square lattice. (b). Schematics for atomic distortions (red arrows) along [1 0 0] and [0 1 0] directions at the phase boundary τ (dashed line). Squares γ and rectangles δ correspond to the in-plane imprints of the rotated tetragonal and orthorhombic lattices.

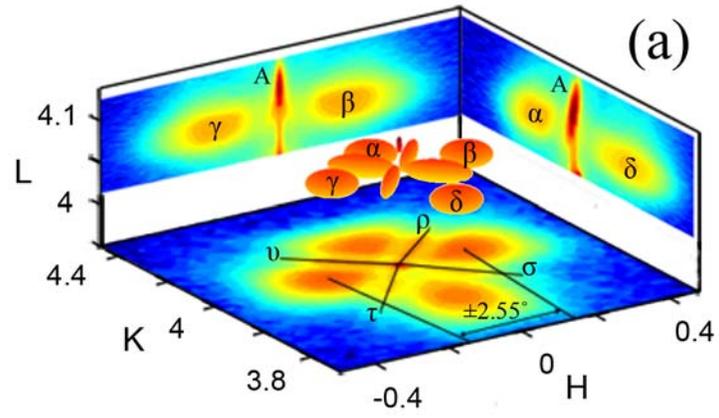
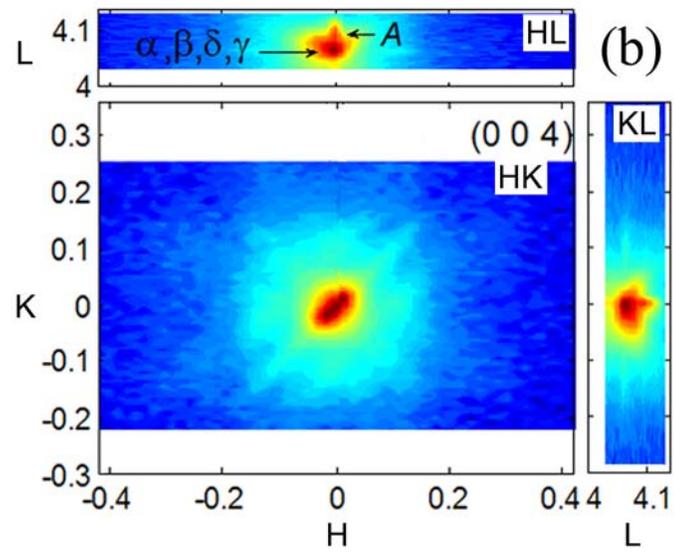

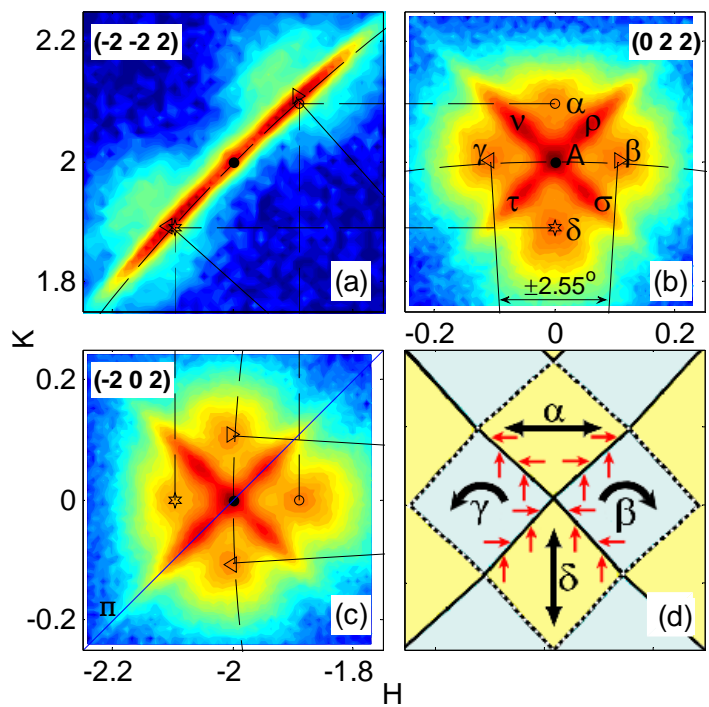

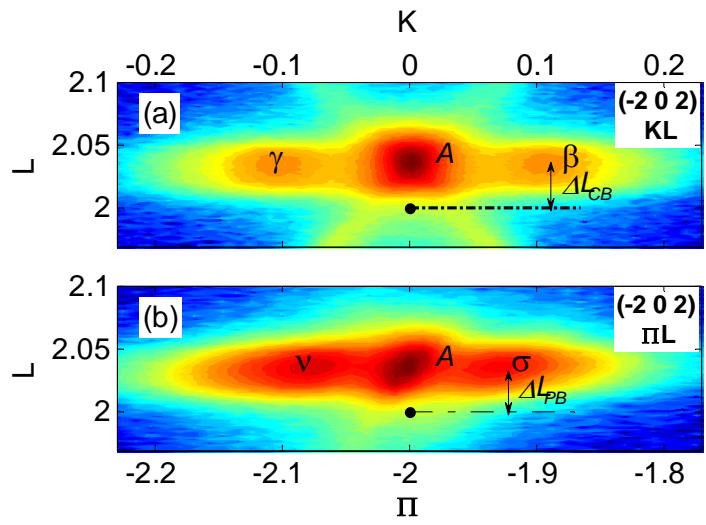

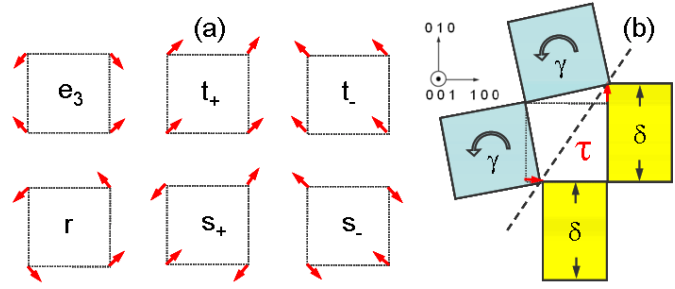